\documentclass[twocolumn,showpacs,showkeys,preprintnumbers,amsmath,amssymb,superscriptaddress,prb]{revtex4}
\usepackage{graphicx,tabularx}
\usepackage{amssymb}
\usepackage[mathcal]{euscript}
\newcommand{\be}{\textrm{B\lowercase{E}(BH}_4\textrm{)}_2}

\newcommand{\beh}{\textrm{B\lowercase{e}H}_2}

\begin{document}
\title{A first-principles study of the electronic structure and stability of Be(BH$_4$)$_2$}
\author{M. J. van Setten}
\affiliation{Electronic Structure of Materials, Institute for Molecules and Materials, Faculty
of Science, Radboud University Nijmegen, Toernooiveld 1, 6525 ED Nijmegen, The Netherlands}
\author{G. A. de Wijs}
\affiliation{Electronic Structure of Materials, Institute for Molecules and Materials, Faculty
of Science, Radboud University Nijmegen, Toernooiveld 1, 6525 ED Nijmegen, The Netherlands}
\author{G. Brocks}
\affiliation{Computational Materials Science, Faculty of Science and Technology and MESA+ Institute
for Nanotechnology, University of Twente, P.O. Box 217, 7500 AE Enschede, The Netherlands}
\date{\today}



\begin{abstract}
Alanates and boranates are studied intensively because of their
potential use as hydrogen storage materials. In this paper we
present a first-principles study of the electronic structure
and the energetics of beryllium boranate, Be(BH$_4$)$_2$. From
total energy calculations we show that - in contrast to the
other boranates and alanates - hydrogen desorption directly to
the elements is likely, and is at least competitive with
desorption to the elemental hydride (BeH$_2$). The formation
enthalpy of Be(BH$_4$)$_2$ is only $-0.12$ eV/H$_2$ (at
$T=0$K). This low value can be rationalized by the
participation of all atoms in the covalent bonding, in contrast
to the ionic bonding observed in other boranates. From
calculations of thermodynamic properties at finite temperature
we estimate a decomposition temperature of 162~K at a pressure
of 1~bar.
\end{abstract}

\maketitle
\section{Introduction}

In the last decade the environmental importance of reducing the
CO$_2$ exhaust has been widely accepted. The use of hydrogen
based fuel cells is an important contribution to achieve this
reduction. One major obstacle for this use is the development
of a method for hydrogen storage with a high gravimetric and
volumetric hydrogen density.\cite{bog_rev}

One way of storing hydrogen is in a (complex) metal hydride.
The ideal hydrogen storage material should have the highest
possible gravimetric hydrogen density. This obviously requires
the use of lightweight materials. Moreover, the formation
energy of the hydride has to be such that it is stable at
atmospheric conditions, yet it has to decompose at a moderate
temperature to release the hydrogen. A further important point
is that the reactions involved in hydrogen de/absorption must
have fast kinetics.

Over the last decade alanates and boranates have been studied
extensively as potential hydrogen storage
materials.\cite{bog_rev,zuttel04} Alanates and boranates
consist of a lattice of metal cations and (AlH$_4)^-$ or
(BH$_4)^-$ complex anions, respectively. Generally these
materials decompose by heating via intermediate complex
hydrides into bulk metals, elemental hydrides and hydrogen gas.
In the last few years the attention has gradually shifted from
alanates towards boranates, because of the high gravimetric
hydrogen density in the latter. Many boranates turn out to be
too stable, however.

In principle a large variety of boranates can be synthesized by
changing the metal cations, which can be used to tune the
formation energy.\cite{nakamori06} So far most effort has been
devoted to the alkali
boranates,\cite{frankcombe05,miwa04,wu03,vajeeston05,au06,wee06,an06}
and more recently to mixtures of alkali
boranates,\cite{miwa05,nakamori06-3} and to the alkaline earth
boranates.\cite{miwa:155122,nakamori06-2,vajeeston06,yu06,chlopek07}
In order to understand the chemical trends we have recently
developed a simple model for the formation energies of these
compounds.\cite{vansetten07jpcc} This model demonstrates that
these boranates are ionic compounds (in the sense discussed
above) and that the difference in their formation energies can
be understood on the basis of the electrostatic (Madelung)
lattice energy. The basic stability of the (BH$_4)^-$ cation is
not affected by substituting one alkali or alkaline earth
cation by another.

The stability of (BH$_4)^-$ may be changed by adding an element
that competes with boron in binding with hydrogen. To
investigate this possibility we study beryllium boranate,
$\be$,\cite{name} in this paper. Establishing the electronic
structure and themodynamic stability of $\be$ will assist in
understanding the chemical and physical trends in alkali,
alkaline earth alanates and boranates.\cite{usability}

We present a first principles study of the electronic structure
and the thermodynamic properties of $\be$. The electronic
structure in relation to the crystal structure is used to
analyze the bonding in $\be$. We calculate total energies and
phonon frequencies of all compounds involved in possible
formation reactions of $\be$. From these data we obtain the
thermodynamic properties at finite temperature.

\section{Computational methods}

First-principles calculations are carried out within the
density functional theory (DFT) approach, applying a
generalized gradient approximation (GGA) for the exchange
correlation functional.\cite{gga} We use a plane wave basis set
and the projector augmented wave (PAW) method,\cite{paw,blo} as
implemented in the Vienna \em Ab initio \em Simulation Package
(VASP),\cite{vasp1,vasp2,vasp3} and apply non-linear core
corrections.\cite{core}

Brillouin zone integrations are performed with a tetrahedron
method\cite{blochl94} for calculating total energies.  A
Gaussian smearing method is used for calculating densities of
states, with a smearing parameter of 0.1~eV. The {\bf k}-point
meshes are such that total energies are converged within
0.1~meV per formula unit. The total energies used in the
calculations of the reaction enthalpies are calculated with a
high plane wave kinetic energy cutoff of 700~eV. By varying the
computational parameters, in particular by trying different PAW
potentials,\cite{pawvariations} we estimate that reaction
enthalpies are converged on a scale of 5~meV.

The atomic positions and lattice parameters are relaxed using a
conjugate gradient algorithm for a range of fixed volumes. The
total energy versus volume curve obtained this way is fitted
with a Murnaghan's equation of state expression, which yields
the ground state volume, the bulk modulus, and its pressure
derivative.\cite{murnaghan44} At the ground state volume we
relaxed the atomic positions and lattice parameters to obtain
the ground state structure. This procedure is followed for all
compounds mentioned in this paper.

To calculate the zero point energies (ZPE) and phonon densities
of state we need the phonon frequencies of all these compounds.
Vibrational frequencies are obtained from the dynamical matrix,
whose matrix elements (the force constants) are calculated
using a finite difference method.\cite{kresse95} The force
constants are calculated from displacements of 0.005~\AA\ in
two opposite directions for each atomic degree of freedom. For
both bulk beryllium and beryllium hydride $2\times2\times2$
supercells give converged ZPEs. One does not need a supercell
to calculate the phonon frequencies of $\be$, since the unit
cell of $\be$ is sufficiently large. For boron we use the
frequencies that have been reported
earlier.\cite{vansetten07-boron}

\section{Crystal structure}

\begin{figure}[!tbp]
\centering
\includegraphics[width=7.0cm,keepaspectratio=true]{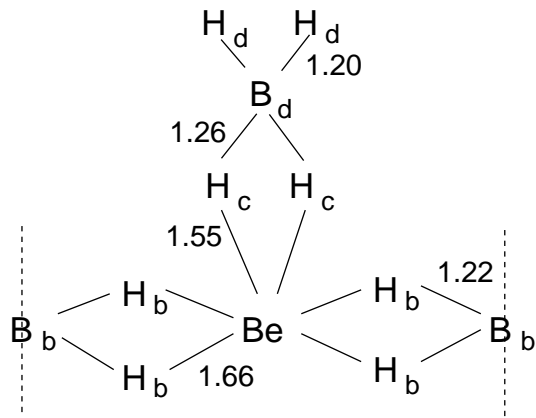}
\caption{Schematic bonding scheme and labeling of the atoms within a polymer chain in $\be$. The
three dimensional structure is given in Fig.~\ref{crystal}. The numbers indicate optimized bond
lengths in \AA.}\label{polschem}
\end{figure}

\begin{figure}[!tbp]
\centering
\includegraphics[width=8.0cm,keepaspectratio=true]{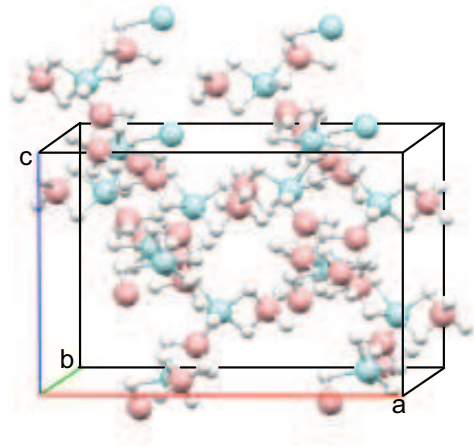}
\caption{(Color online) Crystal structure of $\be$.}\label{crystal}
\end{figure}

$\be$ can be synthesized by the reaction of lithium boranate
and beryllium chloride.\cite{hcp,schlesinger53} Its crystal
structure consists of helical polymers of alternating beryllium
and boron atoms (B$_b$) that are connected via pairs of
hydrogen atoms (H$_b$).\cite{Mary} The polymer building block
is shown schematically in Fig.~\ref{polschem}. A further boron
atom (B$_d$) is attached to each beryllium atom, again via a
pair of hydrogen atoms (H$_c$) and this B$_d$ atom also binds
two ``dangling'' hydrogen atoms (H$_d$). The polymers are
packed in the crystal structure as shown in
Figure~\ref{crystal}. On the basis of this structure one may
expect a strong bonding between the atoms in one polymer chain,
and a much weaker bonding between the polymer chains. The
latter is reflected in the low melting point of $\be$ of
125$^\circ$C.

We relaxed the crystal structure of $\be$ as described in the
previous section, including the cell volume, lattice parameters
and atomic positions, while keeping the experimental
space-group and Wyckoff positions. Except for the cell volume,
the parameters compare well to the experimental
values.\cite{Mary} The calculated cell volume is 15\% larger
than the experimental cell. This indicates that the binding
between the polymer chains is indeed weak and of van der Waals
type. It is well-known that, using the common functionals, DFT
fails to capture van der Waals bonding. However, the total
energy difference between the experimental and calculated cell
volumes is less than 5~meV/H$_2$. This error only has a minor
effect on the relative total energies.

\begin{figure}[!tbp]
\centering
\includegraphics[width=8.0cm,keepaspectratio=true]{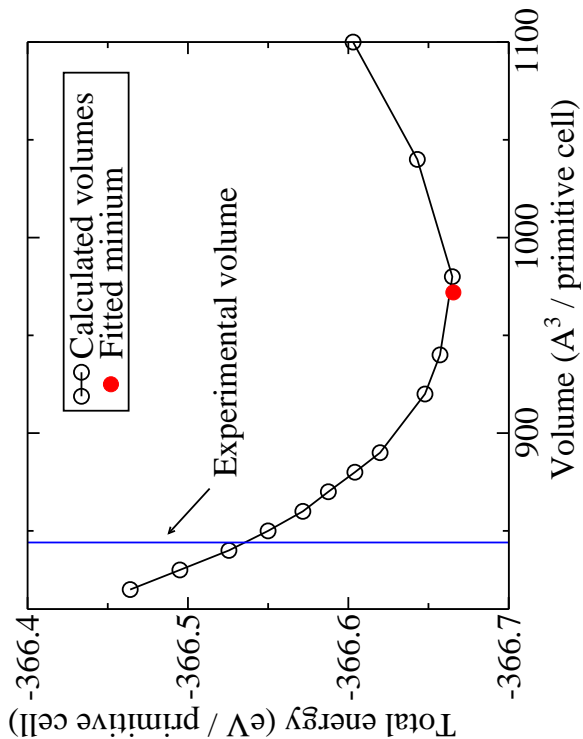}
\caption{Total energy per unit cell of $\be$ as a function of the cell volume.}\label{volen}
\end{figure}

\begin{table}[!tbp]
\caption{Optimized atomic positions of $\be$. The space group
is $I4_1cd$ (110) and all atoms are on Wyckoff positions 16b.
The optimized lattice parameters are $a,b,c = 14.28,14.28,9.54$
\AA. \label{pos}}
\begin{ruledtabular}
\begin{tabular}{lrrr}
atom & x & y & z\\
\hline
Be    & 0.2050 & 0.0992 & 0.0016\\
B$_d$ & 0.1695 & 0.9702 & 0.0068 \\
B$_b$ & 0.1503 & 0.1978 & 0.1237 \\
H$_d$ & 0.0997 & 0.9439 & 0.0653 \\
H$_d$& 0.2183 & 0.9130 & 0.9499 \\
H$_c$ & 0.2189 & 0.0123 & 0.0963 \\
H$_c$ & 0.1453 & 0.0294 & 0.9157 \\
H$_b$ & 0.1083 & 0.1649 & 0.0231 \\
H$_b$ & 0.2281 & 0.1647 & 0.1450 \\
H$_b$ & 0.1611 & 0.2813 & 0.1003 \\
H$_b$ & 0.1027 & 0.1793 & 0.2269 \\
\end{tabular}
\end{ruledtabular}
\end{table}

The optimized B--H and Be--H bond lengths are given in
Fig.~\ref{polschem}. As references, the B--H bond length in a
(BH$_4$)$^-$ anion is 1.21~\AA, whereas a B--H bond length in a
typical three center B--H--B bond is 1.34~\AA.\cite{hcp}
Comparison with these numbers indicates that the B--H bonding
in $\be$ is closer to that in the (BH$_4$)$^-$, although there
is some distortion due to the presence of the Be atom, in
particular on the B$_d$--H$_c$ bond. This could indicate some
competition between B and Be for bonding to hydrogen. For
comparison, the B--H bond lengths in alkali boranates are all
very close to 1.21~\AA. The Be--H bond lengths are still quite
large, however, which indicates a significant ionic
contribution to the bonding.

We have also optimized the structure of $\beh$, see
Table~\ref{posbeh2}. It agrees well with the experimental
structure\cite{smith} and with that obtained in a previous
calculation,\cite{vajeeston04-Be} the largest difference being
that our calculated bulk modulus (21.4~GPa) is $\sim 10$ \%
smaller than that calculated in
Ref.~\onlinecite{vajeeston04-Be} (23.8~GPa). For elemental
boron we use the $\beta$-rhombohedral structure as given in
Ref.~\onlinecite{vansetten07-boron}. For elemental beryllium
(space group P$\overline{3}$m1 (164)) we find lattice
parameters $a = 2.260$~\AA\ and $c = 3.567$~\AA, which compare
well to the experimental values of 2.29~\AA\ and 3.60~\AA,
respectively.\cite{maka}

\begin{table}[!tbp]
\caption{Optimized crystal structure of $\beh$. The space group is Ibam (72) and the optimized
lattice parameters are $a,b,c = 8.967,4.141,7.643$ \AA. The experimental lattice parameters are
$a,b,c = 9.082,4.160,7.707$ \AA.\cite{smith}\label{posbeh2}}
\begin{ruledtabular}
\begin{tabular}{lcrrr}
atom & Wycoff & x & y & z\\
\hline
Be & 4a & 0 & 0 & 0.25 \\
Be & 8j & 0.1677 & 0.1200 & 0 \\
H & 16k & 0.0882 & 0.2241 & 0.1520 \\
H & 8j & 0.3102 & 0.2771 & 0\\
\end{tabular}
\end{ruledtabular}
\end{table}

\section{Electronic structure}

\begin{figure}[!tbp]
\centering
\includegraphics[width=7.0cm,keepaspectratio=true]{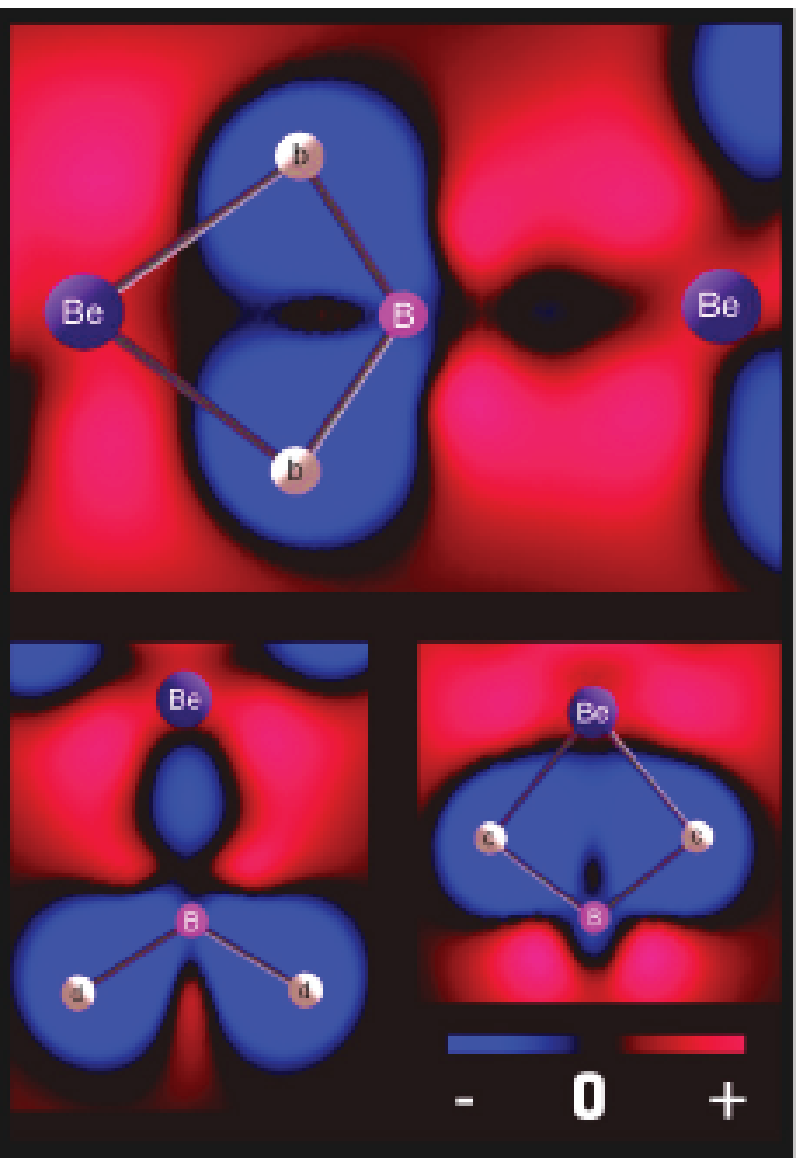}
\caption{(color online) Charge density difference plots of $\be$ with respect to the isolated
atoms. The top picture gives a cut through a plane containing B$_\mathrm{b}$, H$_\mathrm{b}$ and Be
atoms of the polymer backbone, the bottom left picture a cut through the B$_\mathrm{d}$,
H$_\mathrm{c}$ side chain plane, and the bottom right picture a cut through the B$_\mathrm{d}$,
H$_\mathrm{d}$ side chain plane.}\label{difference}
\end{figure}

As discussed above the crystal structure of $\be$ indicates a
weak bonding between polymer chains, and a stronger bonding
within a polymer chain. The charge displacement upon bond
formation can be visualized by plotting the charge density
difference, i.e. the charge density of $\be$ minus that of the
individual isolated atoms. Cuts through the charge density
difference in various planes along a polymer backbone are shown
in Fig.~\ref{difference}. They clearly indicate the formation
of B--H covalent bonds, which are polarized somewhat towards
the H atoms. The character of the Be--H bonds is much less
clear from these plots. In any case these bonds are strongly
polarized in the direction of the H atoms.

\begin{figure}[!tbp]
\centering
\includegraphics[width=8.6cm,keepaspectratio=true]{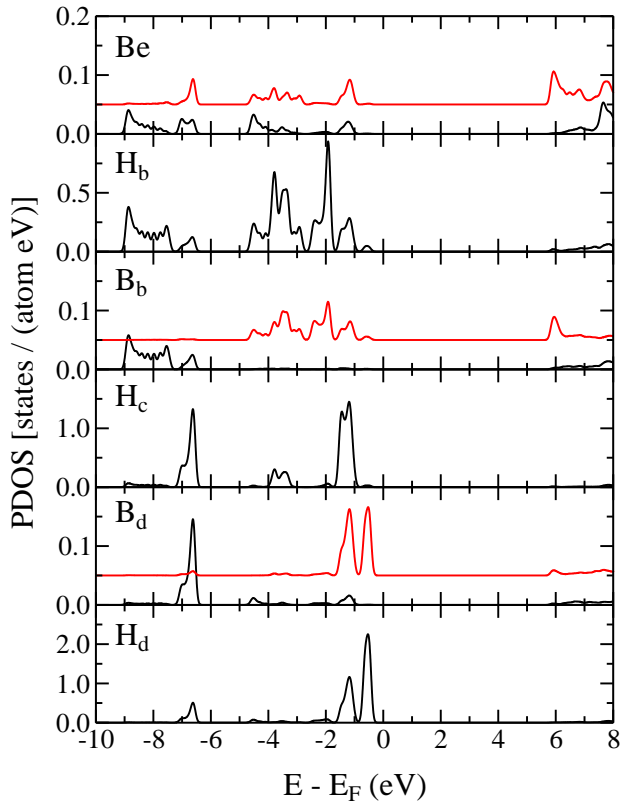}
\caption{(color online) The electronic projected densities of states (PDOS) of $\be$. The Fermi
level, E$_\textrm{F}$, at the top of the valence band is the zero of energy. The upper (red) and
lower curves (black) give projections on $p$ and $s$ atomic states, respectively. Atomic radii of
0.7, 0.5 and 1.1~\AA\ are used for Be, B and H.}\label{pdos}
\end{figure}

The electronic projected density of states (PDOS) of $\be$,
projected on $s,p$ components of the individual atoms, is shown
in Fig.~\ref{pdos}. Tetrahedrally bonded BH$_4^-$ generates a
characteristic pattern in the valence band part of the PDOS,
which is qualitatively similar to that observed for AlH$_4^-$
tetrahedra in the
alanates.\cite{chou,agu,vajeeston04-2,lovvik05,vansetten,vansetten07}
The tetrahedral geometry of BH$_4^-$ results in a splitting
into two valence peaks, the lower one of $s$ (A$_1$) symmetry
and the upper one of $p$ (T$_2$) symmetry, with a relative
weight ratio of 1:3. Projected on atomic states, the $s$-peak
then has contributions from H $s$ and B $s$ orbitals, and the
$p$-peak has contributions from H $s$ and B $p$ orbitals. The
$p$-peak can be split due to symmetry breaking caused by the
crystal field. This is clearly observed in the lowest two
panels of Fig.~\ref{pdos}, showing the PDOS on the
B$_\mathrm{d}$ and H$_\mathrm{d}$ atoms with the $s$ peak at
$\sim -7$ eV, and a $p$ doublet around $\sim -1$ eV. The
splitting between $s$- and $p$-peaks is large ($\sim 6$ eV),
and the crystal field splitting is much smaller ($\sim 1$ eV).

The interaction between the BH$_4$ units in the crystal lattice leads to a broadening of the peaks
due to band formation. The interaction is strongest along the B$_b$(H$_b$)$_4$-Be-B$_b$(H$_b$)$_4$
polymer backbone, see Figs.~\ref{polschem} and \ref{crystal}. This leads to an $s$-type band in the
range $\sim -9$ to $\sim 7.5$ eV, involving contributions from H$_b$, B$_b$, and Be $s$ orbitals,
whose DOS has the characteristic shape of a one-dimensional structure, see the upper three panels of
Fig.~\ref{pdos}. In the range $\sim -5$ to $\sim -2$ eV we find and a set of $p$-type bands. The
band widths are smaller than the $sp$ splitting, but they are not negligible, reflecting the
covalent bonding along the polymer backbone.

\begin{figure}[!tbp]
\centering
\includegraphics[angle=270,width=8.6cm,keepaspectratio=true]{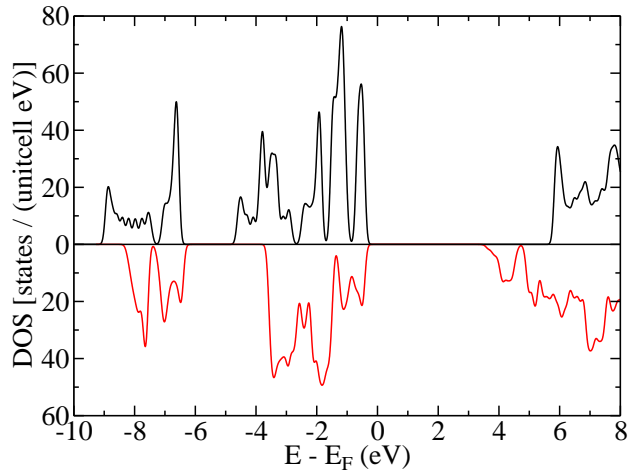}
\caption{(color online) The total electronic densities of states (DOS) of $\be$. The Fermi level,
E$_\textrm{F}$, at the top of the valence band is the zero of energy. In the lower curve the
beryllium atoms are replaced by an homogeneous background with charge $2+$.}\label{dos}
\end{figure}

The involvement of the Be atoms can be clarified by calculating the DOS for a $\be$ structure in
which the Be atoms are replaced by a homogeneous background with charge $2+$. The result is shown
in Fig~\ref{dos}. The $s$ and $p$ valence bands discussed above disappear and are replaced by much
narrower peaks that reflect electron localization on BH$_4^-$ ions in this artificial structure. In
other words, the Be atoms in $\be$ are involved in the covalent bonding. This is in contrast to
alkali or alkaline earth boranates and alanates, where the DOS changes little if the cations are
replaced by a background charge. The bonding in the latter compounds can be described as an ionic
bonding between BH$_4^-$ or AlH$_4^-$ anions and M$^+$ (akali) or M$^{\prime 2+}$ (alkaline earth)
cations.\cite{vansetten07}

\section{Reaction enthalpies}

For light elements such as hydrogen, beryllium and boron the quantum character of their atomic
vibrations is important. This leads to vibrational energies at zero temperature that are not
negligable. For each compound involved in the reaction we calculate its zero point vibrational
energy (ZPVE) from the frequencies of the vibrational modes in the optimized structure. For
hydrogen molecules also the zero point rotational energy (ZPRE) is not entirely negligable.
Reaction enthalpies $\Delta H$ at $T=0$K are then calculated from
\begin{equation}
\label{enthalpy} \Delta H = \sum_{p}\left(E^\mathrm{tot}_p + E^\mathrm{ZPVE}_p\right) +
E^\mathrm{ZPRE}_{H_2}
 - \sum_{r}\left(E^\mathrm{tot}_r + E^\mathrm{ZPVE}_r\right)
\end{equation}
where $E^\mathrm{tot}_{p/r}$ denotes the total energy of the reaction products $p$ or reactants
$r$, $E^\mathrm{ZPVE}_{p/r}$ are the corresponding ZPVEs, and $E^{ZPRE}_{H_2}$ is the ZPRE of the
hydrogen molecules involved in the reaction.

For the hydrogen molecules we calculate a vibrational frequency of 4356~cm$^{-1}$, in good
agreement with the experimental value of 4401~cm$^{-1}$.\cite{huber} The ZPVE, 0.266~eV, is then
calculated from the energy levels of a Morse potential,
\begin{equation}\label{h}
E(n)=\hbar\omega\left(n+\frac{1}{2}\right) -
\frac{1}{4D_e}\left[\hbar\omega\left(n+\frac{1}{2}\right)\right]^2,
\end{equation}
where $\omega$ is the vibration eigenfrequency and $D_e=4.57~eV$ is the dissociation energy. Assuming that ortho- and para-hydrogen are produced in a proportion of three to one, the average ZPRE of a hydrogen molecule is 0.011~eV,
using the energy levels given in Ref.~\onlinecite{huber}.

The calculated total energies and ZPEs of all compounds involved in the reactions are listed in
Table~\ref{energies}. We consider two possible reaction paths for the formation of $\be$. In the
first path $\be$ is directly formed from the elements.
\begin{equation}\label{r1}
\textrm{Be} + 2\textrm{B} + 4\textrm{H}_2(g) \rightarrow \be.
\end{equation}
The second path involves the formation of an intermediate compound $\beh$.
\begin{equation}\label{r2}
\textrm{Be} + \textrm{H}_2(g) \rightarrow \beh.
\end{equation}
\begin{equation}\label{r3}
\beh + 2\textrm{B} + 3\textrm{H}_2(g) \rightarrow \be.
\end{equation}
The enthalpies of these reaction are calculated using Eq.~(\ref{enthalpy}) and the values given in
Table~\ref{energies}.

\begin{table}[!tbp]
\caption{Total energies (with respect to non spin polarized model atoms), zero point
vibrational energies (ZPVE) and zero point rotational energy (ZPRE) in eV/formula
unit in the relaxed structures.}\label{energies}
\begin{ruledtabular}
\begin{tabular}{lrrr}
       & E$^{\textrm{TOT}}$ & E$^{\textrm{ZPVE}}$ & E$^{\textrm{ZPRE}}$\\
\hline
$\be$  &  $-43.353$ & $2.450$ &\\
$\beh$ &  $-10.797$ & $0.542$ &\\
H$_2$  &   $-6.803$ & $0.266$ & $0.011$\\
B      &   $-6.687$ & $0.126$ &\\
Be     &   $-3.729$ & $0.091$ &\\
\end{tabular}
\end{ruledtabular}
\end{table}

Equation~(\ref{r1}) gives a reaction enthalpy of $-0.39$
eV/H$_2$ if ZPEs are neglected. If ZPEs are included the
reaction enthalpy becomes $-0.12$ eV/H$_2$, which indicated the
importance of ZPE corrections for these lightweight compounds.
In principle these values are in a range that is useful for
hydrogen storage. Using the ionic model of
Ref.~\cite{vansetten07jpcc} gives a reaction enthalpy of
$-0.02$ eV/H$_2$ (neglecting ZPEs). In the previous section we
have already concluded that the bonding in $\be$ is not purely
ionic, i.e. Be$^{2+}$(BH$_4$)$^-_2$. The Be atoms are bonded
(partially) covalently to BH$_4$, which increases the bonding
as compared to the pure ionic picture, resulting in a higher
dehydrogenation enthalpy.

Most alanates and other boranates form a simple alkali / alkaline earth hydride when hydrogen is
released in a first step. The dehydrogenation of this simple hydride then occurs as a separate
second step. Usually the enthalpies are such that only the first step is considered useful for
hydrogen storage. For $\be$ these two steps correspond to the reverse reactions of Eqs.~(\ref{r3})
and (\ref{r2}). The calculated reaction enthalpies of Eqs.~(\ref{r2}) and (\ref{r3}) are $-$0.27
and $-$0.43~eV/H$_2$ without ZPEs, and $-$0.09 and $-$0.13~eV/H$_2$ with ZPE corrections,
respectively.

Note that per H$_2$ $\be$ is slightly more stable than $\beh$.
This would indicate that a one-step reaction directly from the
elements, Eq.~(\ref{r1}), is more favorable than the two-step
reaction via the simple hydride, Eqs.~(\ref{r2}) and
(\ref{r3}). The enthalpy difference however is very small. In
addition kinetic barriers may influence the relative importance
of the two reaction paths.

We will now focus on finite temperature properties. For the solids we calculate the Gibbs free
energy $G(T)$ in the harmonic approximation
\begin{equation}
G(T) = E^{\textrm{tot}} + H^{\textrm{vib}}(T) - TS^{\textrm{vib}}(T)
\end{equation}
with
\begin{equation}\label{hvib}
H^{\textrm{vib}}(T) = \int_0^\infty d\omega g(\omega) \left\{ \frac{1}{2}\hbar \omega + \hbar \omega
n(\omega)  \right\}
\end{equation}
and
\begin{eqnarray}
S^{\textrm{vib}}(T) = k_B \int_0^\infty d\omega g(\omega) \left\{\beta\hbar\omega n(\omega) - \ln\left[1 -
e^{ -\beta\hbar\omega}\right] \right\}
\end{eqnarray}
where $g(\omega)$ is the phonon density of states, $n(\omega) = \left[\exp(\beta\hbar\omega) -1
\right]^{-1}$ is the Bose-Einstein occupation number and $\beta = 1/k_BT$. The first term in the
integral of Eq.~(\ref{hvib}) gives the ZPVE and the second term gives the finite temperature
contribution. Note that we neglect the $PV$ term (i.e. the distinction between energy and
enthalpy), which is a good approximation for solids.

\begin{figure}[!tbp]
\centering
\includegraphics[angle=270,width=8.6cm,keepaspectratio=true]{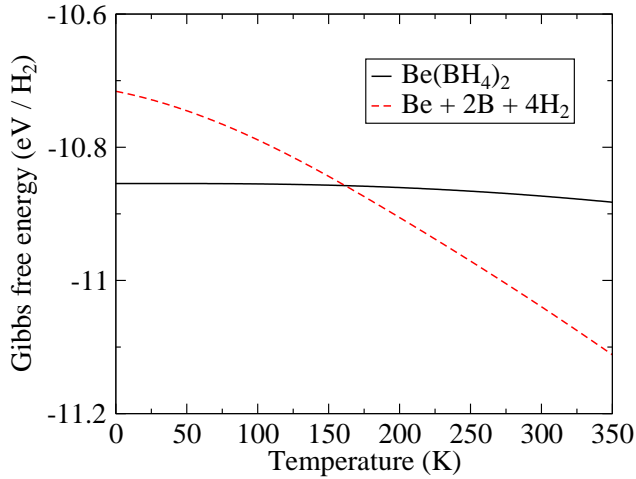}
\caption{(color online) Gibbs free energy of $\be$ and the constituting elements.}\label{gibbs}
\end{figure}

For the Gibbs free energy, the enthalpy and the entropy of the
hydrogen gas, we use the values given in
Ref.~\onlinecite{hemmes86}. Fig.~\ref{gibbs} gives the free
energies of $\be$ and the products of the dehydrogenation
reaction, i.e. the left and right hand sides of Eq.~(\ref{r1}),
at the standard pressure of 1~bar. At 162~K the free energy of
the products drops below that of the hydride phase. This
intersection of the two free energy curves defines the
decomposition temperature.

The predicted desorption temperature and equilibrium pressures rely only on the thermodynamics of the reaction. From the fact that experimentally $\be$ seems to be stable at room temperature one may conclude that kinetic barriers
play an important role in stabilizing $\be$. In this respect $\be$ is similar to other boranates
and alanates, where decomposition temperatures are much higher than what is expected on the basis
of thermodynamics and catalysts have to be applied in order to overcome kinetic barriers.

\section{Conclusions}

We use DFT electronic structure calculations at the GGA level to study the crystal structure,
electronic structure and thermodynamics of $\be$. We optimize the atomic positions and lattice
parameters of all compounds involved in possible formation and dehydrogenation reactions. Both the
crystal structure and the electronic structure indicate that the bonding between B and H atoms is
covalent, and that the bonding between Be and H has covalent as well as ionic contributions. The
crystal structure and the electronic density of states give evidence for [-BH$_4$-Be-]$_n$
(helical) polymers.

The enthalpies of possible formation reactions are calculated including zero point energy
corrections. The latter are obtained by the calculating the phonon frequencies of all compounds
involved in the reactions. Since not only hydrogen, but also boron and beryllium are relatively
light elements, the zero points energies are relatively large for these compounds. The enthalpy of
formation of $\be$ from the elements is $-0.39$~eV/H$_2$ without and $-0.12$~eV/H$_2$ with zero
point energy contributions.

$\be$ differs from other boranates and alanates in that its dehydrogenation to the elements is
thermodynamically slightly more favorable than dehydrogenation via the simple hydride $\beh$. In
alkali or alkaline earth boranates and alanates dehydrogenation always occurs via the alkali or
alkaline earth simple hydride. The different behavior of $\be$ is mainly caused by the high
stability of bulk beryllium metal.

$\be$ follows the general trends in the formation energies that have been observed in alkali and
alkaline earth alanates and boranates. Boranates are more stable than the corresponding alanates,
lighter cations give compounds that are more unstable, and alkaline earth compounds are more
unstable than alkali compounds.\cite{vansetten07} Indeed $\be$ is less stable than LiBH$_4$ or
Mg(BH$_4$)$_2$. We have not found mentioning of beryllium alanate in the literature, which might
indicate that this compound would be too unstable.

Using the calculated phonon spectrum we have calculated free energies within the harmonic
approximation to assess thermodynamic properties at finite temperature. We obtain a decomposition
temperature of 162~K at a 1~bar pressure.

\begin{acknowledgments}
The authors wish to thank Prof. Dr. R.A. de Groot for helpful discussions and J.J. Attema for the
use of his imaging software. This work is part of the research programs of `Advanced Chemical
Technologies for Sustainability (ACTS)' and the `Stichting voor Fundamenteel Onderzoek der Materie
(FOM)', both financially supported by the `Nederlandse Organisatie voor Wetenschappelijk Onderzoek
(NWO)'.
\end{acknowledgments}

\bibliography{proefschrift,notes}

\begin{thebibliography}{46}
\bibitem[{usa()}]{usability}
\bibinfo{note}{$\be$ itself, however, will probably never be used as a hydrogen
  storage material in large scale commercial applications because of the
  toxicity of beryllium gas.}
\end{thebibliography}

\end{document}